\documentclass[12pt]{article}
\usepackage{a4wide}
\usepackage{amssymb}
\usepackage{amsmath}
\usepackage{graphicx}
\usepackage{mathdots}
\usepackage{slashed}
\usepackage{verbatim}
\usepackage{cite}
\usepackage{xcolor}
\usepackage[utf8]{inputenc}
\usepackage{mathtools}

\usepackage[normalem]{ulem}

\usepackage{hyperref}
\usepackage{cleveref}

\usepackage{IEEEtrantools}

\def\makeatletter{\catcode`\@=11}
\makeatletter
\def\mathbox#1{\hbox{$\m@th#1$}}%
\def\math@ccstyles#1#2#3#4#5#6#7{{\leavevmode
      \setbox0\mathbox{#6#7}%
      \setbox2\mathbox{#4#5}%
      \dimen@ #3%
      \baselineskip\z@\lineskiplimit#1\lineskip\z@
      \vbox{\ialign{##\crcr
             \hfil \kern #2\box2 \hfil\crcr
             \noalign{\kern\dimen@}%
             \hfil\box0\hfil\crcr}}}}
\def\mathaccstyles{\math@ccstyles\maxdimen}
\def\maththroughstyles{\math@ccstyles{-\maxdimen}}
\def\unity%
 {\maththroughstyles{.45\ht0}\z@\displaystyle {\mathchar"006C}\displaystyle 1}
 
\usepackage{tikz}
\usetikzlibrary{positioning,arrows}
\usetikzlibrary{decorations.pathmorphing,calc}
\usetikzlibrary{decorations.markings}
\newif\ifmirrorsemicircle
\usepackage{float}
\usepackage{ifthen}
\usepackage{enumerate}
\usepackage{amsmath}
\usepackage{amssymb}
\usepackage[mathscr]{eucal}
\usepackage[errorshow]{tracefnt}
\usepackage{amsmath}
\usepackage{slashed}
\usepackage{amssymb}
\usepackage{array}
\usepackage{amsthm}
\usepackage{eucal}
\usepackage{epsf}
\usepackage{epsfig}
\usepackage{psfrag}
\usepackage{multirow}
\usepackage{wasysym} 
\usepackage{tikz-cd} 
\usepgfmodule{shapes}
\usetikzlibrary{backgrounds, arrows,calc,shapes,decorations.pathreplacing, decorations.markings, automata,positioning}
\usepackage{verbatim}

 \numberwithin{equation}{section}
\begin{document}

\begin{flushright}\footnotesize

\texttt{}
\vspace{0.6cm}
\end{flushright}

\mbox{}
\vspace{0truecm}
\linespread{1.1}

\centerline{\LARGE \bf R-symmetries, anomalies and non-invertible defects}
\medskip
\medskip
\centerline{\LARGE \bf from non-BPS branes}
\medskip

\vspace{.5cm}

 \centerline{\LARGE \bf }

\vspace{1.5truecm}

\centerline{
    {\bf Hugo Calvo${}^{a,b}$} \footnote{calvohugo@uniovi.es},
    { \bf Francesco Mignosa${}^{a,b}$} \footnote{francesco.mignosa02@gmail.com}
        {\bf and}
    { \bf Diego Rodriguez-Gomez${}^{a,b}$} \footnote{d.rodriguez.gomez@uniovi.es}}

\vspace{1cm}
\centerline{{\it ${}^a$ Department of Physics, Universidad de Oviedo}} \centerline{{\it C/ Federico Garc\'ia Lorca  18, 33007  Oviedo, Spain}}
\medskip
\centerline{{\it ${}^b$  Instituto Universitario de Ciencias y Tecnolog\'ias Espaciales de Asturias (ICTEA)}}\centerline{{\it C/~de la Independencia 13, 33004 Oviedo, Spain.}}
\vspace{1cm}

\centerline{\bf ABSTRACT}
\medskip 
 
We propose a holographic realization of  symmetry operators for symmetries associated to isometries in terms of non-BPS Kaluza-Klein monopoles. Their existence is supported by dualities, and their action can be inferred by consistency with well-known String Theory constructions. We use our proposal to describe the $U(1)$ superconformal R-symmetry of the Klebanov-Witten 4d $\mathcal{N}=1$ theory dual to Type IIB String Theory on $AdS_5\times T^{1,1}$. We precisely reproduce the expectations from field theory, including the 't Hooft self-anomaly of the R-symmetry and the mixed 't Hooft anomaly with the baryonic symmetry. For a choice of boundary conditions, the R-symmetry becomes non-invertible and the worldvolume action for the non-BPS Kaluza-Klein monopole precisely accounts for this.

\noindent

\newpage

\tableofcontents

\section{Introduction}

The $AdS$/CFT correspondence establishes a map between String Theory on a specific $AdS\times X$ background and a certain Quantum Field Theory (QFT) living on its boundary. The original version of the correspondence states that the global symmetries of the QFT are mapped to gauge symmetries of the bulk String Theory \cite{Witten:1998qj}. In recent years, this was extended and analyzed in the context of the Symmetry Topological Field Theory (SymTFT)  \cite{Freed:2012bs, Gaiotto:2020iye, Apruzzi:2021nmk, Freed:2022qnc} and the Symmetry Theory (SymTh) \cite{Apruzzi:2024htg, Bonetti:2024cjk}. In particular, operators of the QFT charged under the global symmetry correspond, in String Theory, to wrapped BPS branes ending on the boundary of $AdS$. It is then natural to search for the holographic description of the associated symmetry operators. For discrete symmetries, after reduction over the internal space, it has been argued that BPS branes pushed to the $AdS$ boundary can link appropriately with the charged operators, thus implementing the corresponding global symmetry close to the boundary \cite{OBtalks,Apruzzi:2022rei,GarciaEtxebarria:2022vzq,Bergman:2022otk,Heckman:2022muc,Apruzzi:2023uma, Bah:2023ymy}. This description cannot be extended to continuous symmetries. In fact, the String Theory description of continuous symmetry operators seems \textit{a priori} difficult to imagine given that these are labelled by a continuous parameter which appears hard to obtain from the worldvolume action of BPS branes. A resolution to this puzzle was put forward in \cite{Bergman:2024aly}, which suggested that non-BPS branes \cite{Sen:1998tt, Sen:1999md, Sen:1999mg, Sen:2003tm} in the tachyon vacuum are the right objects to describe symmetry operators for continuous symmetries.\footnote{See \cite{Gutperle:2001mb,  Cvetic:2023plv} for an alternative description of these operators in terms of flux branes.} In particular, it was argued that they correctly link with the BPS branes describing charged operators\footnote{It should be stressed that non-BPS branes corresponding to symmetry operators link with BPS branes corresponding to charged operators in the full 10d $AdS \times X$ space.} and moreover they carry the aforementioned continuous parameter. The construction was successfully put to work  \cite{Bergman:2024aly, Calvo:2025kjh} in several examples such as the Klebanov-Witten theory \cite{Klebanov:1998hh}, where it was shown that a non-BPS D4-brane was able to implement the baryonic symmetry of the dual QFT. 

An important part of the proposal in \cite{Bergman:2024aly} is that non-BPS branes naturally link with the BPS branes holographically dual to charged operators, which are associated to symmetries captured by fluctuations of the RR or NSNS gauge fields (such as the baryonic symmetry of the Klebanov-Witten theory). As a consequence, the description of holographic symmetry operators in terms of non-BPS branes did not include symmetries associated to isometries of the gravitational background --such as R-symmetries-- whose holographic description appears more cumbersome. The purpose of this paper is to remedy this. To add further complications, these symmetries are often non-abelian (\textit{e.g.} the $SO(6)$ R-symmetry of $\mathcal{N}=4$ SYM, or the $SU(2)\times SU(2)$ mesonic symmetry of the Klebanov-Witten theory). To ease this last point, we concentrate on abelian symmetries focusing in particular on the case of the $U(1)$ R-symmetry of the Klebanov-Witten theory. Our goal in this paper is to propose a candidate for a holographic description of R-symmetry operators. Given that the R-symmetry and the baryonic symmetry have mixed 't Hooft anomalies, by identifying the right holographic symmetry operators with their expected anomalies we also provide a highly non-trivial check of the proposed description of symmetry operators in terms of non-BPS branes.

In this paper we propose that the objects which holographically describe abelian symmetries associated to isometries are non-BPS Kaluza-Klein (KK) monopoles.\footnote{See \cite{Lawrie:2023tdz, Waddleton:2024iiv, Cvetic:2025kdn} for an alternative proposal.} We argue that their existence is required by duality arguments and that they must decay into BPS KK monopoles, which have been thoroughly studied in String Theory \cite{Eyras:1998hn}. Consistency then allows to determine their worldvolume action. Applying this to the particular case of the Klebanov-Witten theory, we show that the worldvolume action of non-BPS KK monopoles precisely reproduces the expected behavior of the $U(1)_R$ symmetry operators including their 't Hooft anomalies. Moreover, we complete the picture by also showing that the non-BPS D4-branes introduced in  \cite{Bergman:2024aly} as holographic duals to the baryonic symmetry operators also capture the correct 't Hooft anomalies. An important point in the construction is that, due to the anomalies, for a certain choice of boundary conditions, the branes can implement non-invertible continuous symmetries for which our String Theory construction picks a definite realization similar to \cite{GarciaEtxebarria:2022jky}.

The content of the paper is as follows. In section \ref{Section1}, we show how non-BPS monopoles couple with the gauge fields associated with the isometry of their Taub-NUT direction. We also see how the corresponding charged operators can be identified in terms of a gravitational wave \cite{Bergshoeff:1994dg, Janssen:2002vb} propagating in the Taub-NUT direction. Similarly, a BPS KK-monopole represents a source for the dual gauge field, whose flux can be measured by a non-BPS gravitational wave.  In section \ref{Section3}, we focus on a specific example: the Klebanov-Witten theory. We start analyzing the symmetries and anomalies of this theory and of the theory obtained from gauging the $U(1)_B$ baryonic symmetry. We then consider the reduction of String Theory on $AdS_5\times T^{1,1}$. The corresponding theory in $AdS$ describes the SymTh of the Klebanov-Witten theory. We analyze its charged and topological operators, their correlation functions, and the constraint on the possible boundary conditions, matching these properties with the Klebanov-Witten theory and its gauged version. In section \ref{Section4}, we apply the procedure of section \ref{Section1}, calculating explicitly the reduction of the worlvolume theory of the non-BPS KK monopole on the base $S^2\times S^2$ of $T^{1,1}$ and matching it with the symmetry operators implementing the R-symmetry in the SymTh. We then similarly determine all the other symmetry and charged operators in terms of non-BPS/BPS branes reduced on cycles of the internal geometry. In \cref{Conclusions} we conclude and give some outlook of this work. Finally, we compile in Appendix \ref{AppendixConifold} several details and useful formulae concerning the geometry of $T^{1,1}$.

\section{Symmetry operators for isometries}\label{Section1}

As it is well-known, isometries of the gravitational background correspond to global symmetries of the dual holographic QFT, including in particular its R-symmetry. In the following, we will mostly focus on $\mathcal{N}=1$ 4d Superconformal Field Theories (SCFTs) holographically dual to Type IIB String Theory backgrounds of the form $AdS_5\times X$ with $X$ a Sasaki-Einstein space (see \textit{e.g.} \cite{Acharya:1998db,Morrison:1998cs,Benvenuti:2004dy}), of which the Klebanov-Witten theory \cite{Klebanov:1998hh} --for which $X=T^{1,1}$-- is a particular example. Holographically, the space $X$ is $U(1)$ bundle over a 4-dimensional base $B$ and its metric is of the form

\begin{equation}
ds_X^2=g_{\phi\phi}(d\phi+\xi)^2+ds_B^2\,,
\end{equation}
with $\xi$ a $B$-valued 1-form and $\phi$ is the fiber direction --which corresponds to a Reeb vector-- and $g_{\phi\phi}$ is $B$-valued. 

On general grounds, these SCFTs enjoy a $U(1)$ R-symmetry whose current sits in the same superconformal multiplet as their energy-momentum tensor. Holographically, the R-symmetry is associated to the isometry along the fiber $\phi$ \cite{Berenstein:2002ke}.  Gauging this introduces a field $A$

\begin{equation}
ds^2=ds_{AdS_5}^2+g_{\phi\phi}(d\phi+\xi+ A)^2+ds_B^2\,,
\end{equation}
which is $AdS_5$-valued and it is associated with the diffeomorphism $(A,\,\phi)\rightarrow (A-d\lambda,\,\phi+\lambda)$ with $\lambda$ a scalar function in $AdS_5$. Note that, in addition to $A$, generically, the full KK fluctuation will excite scalars regulating the volumes and shapes of the internal cycles. Since we are interested on the near-boundary region, where we expect the scalars to die off, we can focus only on the gauge field sector.

Before dealing with the symmetry operators, let us discuss the operators charged under this isometry. In QFT the typical operators charged under the R-symmetry are mesons,\footnote{Baryons are also charged, but we will come back to those later.} which are effectively described as supergravity KK fluctuations of the background geometry. The associated particle is a gravitational wave (GW) whose direction of propagation encodes the charges of the meson.\footnote{This point of view has proved useful in the past. For instance, it allowed \cite{Janssen:2002cf,Janssen:2003ri} to microscopically describe giant gravitons, which are nothing but large mesons.} A worldvolume action for GW was constructed in \cite{Janssen:2002vb} using the well-known fact that GWs and fundamental strings are dual under T-duality along the worldvolume of the string, which then becomes the direction of propagation of the GW. In a specific gauge where the modes of the string $X^\alpha(\sigma, \tau)$ can be separated as 
 
\begin{equation}
X^\alpha(\sigma, \tau)= (X^M(\tau), X^\phi(\sigma))=(X^M(\tau), m\sigma)\,,
\end{equation}
where $m$ is the number of times the string winds the compact direction, the Nambu-Goto action of the string is mapped to the action of the GW which reads

\begin{equation}
S_{\text{GW}}= -m T_{\text{GW}}\int_{M_1} d\tau \kappa^{-1}\sqrt{|\partial_\tau X^M \partial_\tau X^N \mathcal{G}_{MN}|}+ mT_{\text{GW}}\int_{M_1} \imath_{\kappa}\tilde{B}_2\,, \qquad T_{\text{GW}}=2\pi\, T_{\text{F1}}=1
\end{equation}
where $\kappa=\delta^{\alpha}_\phi\partial_\alpha$, so $\kappa^2=-\kappa_\alpha \kappa^\alpha=\tilde{g}_{\phi\phi}$ and $\kappa_M= \tilde{g}_{M\phi}, \,\,\, \kappa_\phi= \tilde{g}_{\phi\phi}$ and $\tilde{B}_2$ is the NSNS 2-form coupled to the string.\footnote{As we will often need to consider T-dual frames, in the following we will denote by $\tilde{\cdot}$ quantities in the T-dual background. In this respect, it is important to remember that to apply Buscher's T-duality rules, the dualized direction must have period $2\pi$.} Applying this to our case, mesons charged under the R-symmetry should correspond to GW propagating along $\xi$ in our background. From Buscher's rules it then follows that

\begin{equation}
\tilde{B}_2= -(\xi+ A)\wedge d\phi\,.
\end{equation}
Hence, focusing on the WZ term, for a probe gravitational wave moving along $\phi$, we have (here we are setting $g_s$ and $\alpha'$ to 1)

\begin{equation}
\label{SGW}
S_{\text{GW}}^{\text{WZ}}= -\int_{M_1} A\,.
\end{equation}
Thus the worldvolume theory of the gravitational wave has the form of a Wilson line of $A$, representing the holographic realization of the operator charged under the R-symmetry when ending at the boundary of $AdS$.

Having identified the charged operators, let us turn to the symmetry operators. Given that these must capture the flux sourced by the charged operator, which in this case is purely gravitational, it seems clear that no non-BPS nor BPS brane can do the job. Nevertheless, String Theory contains objects which are sensible to background isometries: KK monopoles, which come automatically equipped with an isometric Taub-NUT direction. It is thus natural to conjecture that a non-BPS version of the KK monopole has to be identified with the symmetry operator linking with the GW. 

The existence of non-BPS KK monopoles can be predicted by S- and T-duality as follows. Starting in Type IIB, we have non-BPS D6 branes. A tachyon kink on their worldvolume gives rise to a localize BPS D5 brane \cite{Sen:2003tm}. By S-duality, we must then have non-BPS NS6 branes, whose tachyon kink realizes a BPS NS5$_B$ brane \cite{Intriligator:2000pk}. Upon T-duality along one of its transverse directions, the non-BPS NS6 gets mapped into another object in type IIA String Theory, which we denote by $\overline{\text{KK}}_A$. Since T-duality along a transverse direction of NS5$_B$ gives Type IIA KK$_A$ monopoles, consistency requires that $\overline{\text{KK}}_A$ decays into KK$_A$. Hence we see that $\overline{\text{KK}}_A$ plays the role of a non-BPS brane for a KK monopole, namely it describes a non-BPS KK-monopole. Consistency with T-duality along a worldvolume direction then requires the existence in IIB of a non-BPS KK$_B$-monopole supporting a tachyon realizing the decay into a type IIB BPS ${\rm KK}_B$ monopole. This is the object we are interested in. We summarize in \cref{dualities} our chain of dualities.

\begin{figure}
\begin{equation}
\begin{array}{ |c|  l  c l |c| l c l |c| l c l |c|}
\overline{\text{D6}} & & & & \overline{\text{NS6}}_B & & & & \boxed{\overline{\text{KK}}_A} & & & & \boxed{\overline{\text{KK}}_B} \\
 & & & & & & & &  & & & & \\
 \Downarrow & & \xlongrightarrow{S} & &  \Downarrow & & \xlongrightarrow{T_{\perp}} &  &  \Downarrow  & &\xlongrightarrow{T_{\|}} & &  \Downarrow \\
   & & & & & & & &  & & & & \\
\text{D5} & & & & \text{NS5}_B & & & & \text{KK}_A & & & & \text{KK}_B
\end{array}
\end{equation}
\caption{Sequence of dualities predicting the boxed objects. Vertical arrows stand for decay through a tachyon kink.}
\label{dualities}
\end{figure}

To see that the $\overline{\text{KK}}_B$ couples to the Hodge dual field strength of the R-symmetry field, let us first look to the worldvolume of a BPS KK$_B$ monopole. This was constructed in \cite{Eyras:1998hn} and, on top of the transverse directions it wraps, it has an associated Taub-NUT direction related to an isometry of the internal space. Its WZ action reads \cite{Eyras:1998hn, Eyras:1998rf} 

\begin{equation}
S_{\text{KK}}^{\text{WZ}}=- 2\pi T_{\text{KK}_B}\int_{M_6} \mathcal{K}_6\,,\qquad  T_{\text{KK}_B}= \frac{1}{(2\pi)^5}\,,
\end{equation}
where $\mathcal{K}_6$ is a curvature that can be expressed in terms of the background gauge fields and worldvolume ones, which full expression can be found in \cite{Eyras:1998hn}. The term of interest is the monopole coupling within $\mathcal{K}_6$

\begin{equation}
\mathcal{K}_6=\frac{\imath_{\kappa}\mathcal{N}}{2\pi}+\cdots\,,
\end{equation}
where $\mathcal{N}$ is a 7-form potential to which KK$_B$ couples and is defined as  $\imath_{\kappa}\mathcal{N}=\tilde{B}_6$ in the corresponding T-dual type IIA background. So, analogously to standard non-BPS branes, the non-BPS KK$_B$ monopole can be constructed as an object suffering an instability driven by a real tachyon $T$ with WZ term

\begin{equation}
S_{\text{KK}_B}^{\text{WZ}}=-T_{\overline{\text{KK}}_B} \int_{M_7} W(T)\,dT\wedge \imath_{\kappa}\,\mathcal{N}\,,
\end{equation}
with $W(T)$ has the same properties as the tachyon potential $V(T)$ \cite{Bergman:2024aly} and $T_{\overline{\text{KK}}_B}=\sqrt{2}T_{\text{KK}_B}$. 
Integrating by parts and going to the tachyon vacuum the only surviving part of the non-BPS KK monopole worldvolume action is the WZ above, so

\begin{equation}
S_{\overline{\text{KK}}_B}^{\text{WZ}}= \alpha\,T_{\overline{\text{KK}}_B} \int_{M_7} d(\imath_{\kappa}\,\mathcal{N})\,,
\end{equation}
where $\alpha$ represents a $U(1)$ parameter associated with the integration by parts \cite{Bergman:2024aly}.

Let us finally write $d(\imath_{\kappa}\,\mathcal{N})$ in terms of type IIB fields and see where the isometry field enters in the game. T-duality along $\phi$ produces

\begin{align*}
&d\tilde{s}^2= ds_{AdS_5}^2+ds_B^2 +g_{\phi\phi}^{-1}d\phi^2\,,\qquad \tilde{B}_2= -(\xi+ A)\wedge d\phi\,,\\
&\tilde{C}_3=- \imath_{\kappa} C_4\,, \qquad  e^{-2\tilde{\Phi}}=e^{-2\Phi}g_{\phi\phi}\,.
\end{align*}
Focusing on the isometry field, we see that 

\begin{equation}
\tilde{H}_3=-dA\wedge d\phi\,.
\end{equation}
Its Hodge dual with respect to the type IIA metric is

\begin{equation}
\label{deltaH3}
\tilde{\star} \tilde{H}_3=-\star_{AdS_5} dA\wedge {\rm Vol}(B)\,,
\end{equation}
where $\star_{AdS_5}$ is the 5d $AdS_5$ Hodge-star and ${\rm Vol}(B)$ is the volume form on the space $B$. Recalling that $d(\imath_{\kappa}\mathcal{N})=d\tilde{B}_6$ and the equations of motion of $\tilde{H}_7= e^{-2\tilde{\Phi}} \tilde{\star} \tilde{H}_3$ in Type IIA read 

\begin{equation}
d\tilde{H}_7= \tilde{F}_2 \wedge \tilde{F}_6 -\frac{1}{2} \tilde{F}_4\wedge \tilde{F}_4\,,
\end{equation}
we see that $d(\imath_{\kappa}\mathcal{N})\sim  e^{-2\tilde{\Phi}}\,\tilde{\star}H_3$. Note that strictly speaking there are additional contributions of the RR fields that can be relevant in the definition of $\imath_{\kappa}\mathcal{N}$. We will come back on them in section \ref{Section3}. Then, comparing to eq. \eqref{deltaH3}, we see that the non-BPS KK monopole wrapped on a 3-manifold $M_3$ in $AdS_5$ and $B$ couples with the Hodge dual R-symmetry field strength as

\begin{equation}
S_{\overline{\text{KK}}_B}^{\text{WZ}}= -\alpha\, T_{\overline{\text{KK}}_B}\,I(B)\int_{M_3} \star dA\,,
\end{equation}
with $I(B)= \int_{B} e^{-2\Phi} g_{\phi\phi} \text{Vol}(B)$. Thus the KK monopole behaves just as expected for a symmetry operator.

\subsection{Magnetic states}

As described, the GW plays the role of the Wilson line for the isometry symmetry. In order to find the 't Hooft line, we consider a BPS KK$_B$ monopole wrapping the internal space with isometric direction $\phi$ and spanning a 2d manifold in $AdS_5$. An analogous computation to that for the Wilson line shows that this object sources $\tilde{A}$ defined as $d\tilde{A}=\star dA$. Hence, it plays the role of a 't Hooft line of the R-symmetry field. In turn, the object linking with it is the non-BPS analog of a gravitational wave. Again its existence in type IIB is expected from the existence of non-BPS D2 branes. Indeed, S-duality maps these branes to non-BPS NS2 branes \cite{Intriligator:2000pk} which decay into F1 strings. Similarly, we expect these NS2 to be present in type IIA. As it is well-known, if we T-dualize along the  worldvolume direction a IIA F1 strings, we find a IIB GW which can be regarded as a ''rigid F1" of IIB, that is, a graviton. Then, T-dualizing the NS2, we would expect a rigid non-BPS NS2 (NS2$_R$) as a result, which decays in IIB into the GW. We summarize in \cref{dualities2} the chain of dualities.

\begin{figure}
\begin{equation}
\begin{array}{ |c|  l  c l |c| l c l |c| l c l |c|}
\overline{\text{D2}} & & & & \overline{\text{NS2}}_B & & & & \overline{\text{NS2}}_A & & & & \boxed{\overline{\text{NS2}}_R} \\
 & & & & & & & &  & & & & \\
 \Downarrow & & \xlongrightarrow{S} & &  \Downarrow & & \xlongrightarrow{T_{\perp}} &  &  \Downarrow  & &\xlongrightarrow{T_{\|}} & &  \Downarrow \\
   & & & & & & & &  & & & & \\
\text{D1} & & & & \text{F1}_B & & & & \text{F1}_A & & & & \text{GW}_B
\end{array}
\end{equation}
\caption{Sequence of dualities predicting the boxed objects. Vertical arrows stand for decay through a tachyon kink.}
\label{dualities2}
\end{figure}

Given eq. \eqref{SGW}, this suggests that the WZ of such NS2$_R$ must be

\begin{equation}
\label{SNS2}
S_{\text{NS2}_R}^{\text{WZ}}=-\alpha T_{\text{NS2}_R} \int_{M_2} dA=-\alpha  T_{\text{NS2}_R} \int_{M_2} \star d\tilde{A},\qquad T_{\text{NS2}_R}=\sqrt{2} T_{\text{GW}}\,.
\end{equation}
This implements the magnetic dual symmetry associated with the R-symmetry gauge field. 

\section{SymTh of Klebanov-Witten theory}\label{Section3}

The Klebanov-Witten theory is a 4d ${\cal N}=1$ supersymmetric gauge theory with gauge group $SU(N)\times SU(N)$ and chiral multiplets $A_i, \,\, B_i$ in the $({\bf N},\bar{\bf N})$ and $(\bar{\bf N},{\bf N})$ representations of the gauge groups respectively \cite{Klebanov:1998hh}. The theory has a baryonic $U(1)_B$ global symmetry under which the fields $A_i$ and $B_i$ carry opposite charges. While the usual mesons are neutral under this symmetry, one can construct determinant-like gauge-invariant operators which, in the appropriate normalization, carry charge $N$ under the baryonic symmetry \cite{Gubser:1998fp}. The theory also admits a non-anomalous R-symmetry, under which the chiral multiplets $A_i, \, \, B_i$ have charge $\frac{1}{2}$ and the gauginos of the two nodes have R-charge $1$.  

The baryonic and R-symmetry do not suffer any ABJ anomaly. On the other hand, they have mixed 't Hooft anomalies, which in the obvious notation read

\begin{align}\label{anomalies}
&\mathcal{A}_{RBB}\equiv\sum_i q_R^i (q_B^i)^2= -2N^2, \\
&\mathcal{A}_{RRR}\equiv\sum_i (q_R^i)^3= \frac{3}{2}N^2-2\,,
\end{align}
where the sum runs over all charged fields (bifundamentals and the gauginos).
The presence of the previous 't Hooft anomalies does not allow to gauge the R-symmetry. In turn, gauging the baryonic symmetry is allowed. Upon doing that, the bifundamentals fall into a $({\bf N},\bar{\bf N}, \pm 1)$ representation under the new $SU(N)\times SU(N)\times U(1)_B$ gauge group. The R-symmetry current $J_R$ is no longer conserved due to the ABJ anomaly with the new baryonic gauge field $a_B$

\begin{equation}
d\star J_R=\frac{N^2}{4\pi^2} F_B\wedge F_B, \qquad F_B= da_B\,.
\end{equation}

This implies that (on top of a remnant discrete subgroup acting invertibly) the R-symmetry becomes non-invertible \cite{Choi:2022jqy, Cordova:2022ieu, Karasik:2022kkq, Roumpedakis:2022aik, Arbalestrier:2024oqg} and the original symmetry operators

\begin{equation}
D_R^{(3)}(M_3)=e^{i \alpha \oint_{M_3} \star J_R}\,,
\end{equation}
need to be dressed by a TQFT.  This aspect will be studied in detail in section \ref{Linkingandbcs} in the context of the SymTh.

On top, the theory possesses a 1-form symmetry coming from the diagonal subgroup of the two $SU(N)$ centers. Indeed, lines in the $({\bf N},\bar{\bf N})$ can be screened by a single bifundamental $A_i$ or $B_i$, while the lines $({\bf N},{\bf N})$ are unscreenable ($N$ of these lines can be screened by a combination of the two gluons).  The gauging of the baryonic symmetry introduces an additional 1-form symmetry associated with the Bianchi identity of the gauged baryonic field $a_B$. The charged objects are 't Hooft lines. Note that the fate of the would-be electric 1-form symmetry is identical to that of the would-be antidiagonal $SU(N)\times SU(N)$. One way to see this is that baryonic Wilson lines would be screened by baryons, which are determinants of bifudamental fields and hence can screen themeselves lines in the $(\mathbf{N},\bar{\mathbf{N}})$ representation of the $SU(N)\times SU(N)$. We postpone for further work the detailed study of 1-form symmetries.

\subsection{SymTh from the holographic dual background}

Having analyzed the charged and symmetry operators of the Klebanov-Witten theory, we would like to see how these are realized as operators in holography. To do so, we consider the holographic dual of Klebanov-Witten, which is Type IIB String Theory on $AdS_5 \times T^{1,1}$ with $N$ units of $\tilde{F}_5$ flux. Topologically, the base of the conifold is $T^{1,1}\sim S^2\times S^3$. In the following, we will focus on gauge field fluctuation above the background dual to the baryonic and R-symmetry currents. 

The complete reduction of IIB Supergravity on the $T^{1,1}$ was performed in \cite{Cassani:2010na}. The Klebanov-Witten vacuum corresponds to 

\begin{equation}
w=0\,,\qquad u=v=\frac{1}{4}\log\frac{k}{2}\,,
\end{equation}
where in our conventions the $AdS$ radius is related to $k$ as $k=2L^4$ with

\begin{equation}
L^4=\frac{27}{4}\pi N\,,
\end{equation}
and the metric reads (see Appendix \ref{AppendixConifold})
\begin{equation}
ds^2=\left(\frac{2}{k}\right)^{\frac{5}{6}}\,ds_{AdS_5}^2+\left(\frac{k}{2}\right)^{\frac{1}{2}}ds_{T^{1,1}}\,.
\end{equation}

As anticipated, we are mostly interested on the gauge field sector $\{A,\,a_1^J,\,a_1^{\Phi}\}$ in the reduction of \cite{Cassani:2010na} (even though for reasons which will become momentarily apparent, we will keep the extra scalar $a$). While $A$ appears both in the metric and the expansion on $F_5$, the fields $a_1^J, \, a_1^\Phi$ just appear in the 5-form field strength. In particular, the $F_5$ reduction is

\begin{eqnarray}
\label{F5}
F_5&=&(f_3^{\Phi}+f_2^{\Phi}\wedge A)\wedge \Phi+(f_3^J+f_2^J\wedge A)\wedge J+f_2^{\Phi}\wedge \Phi\wedge \eta+f_2^J\wedge J\wedge \eta \nonumber \\ && +k\,J\wedge J\wedge (\eta+A)+f_1\,J\wedge J+f_4\wedge (\eta+A)\,,
\end{eqnarray}
where
\begin{eqnarray}
&& f_2^J=da_1^J\,,\qquad f_3^J=-\left(\frac{2}{k}\right)^{\frac{2}{3}}\star da_1^J \,,\\
&& f_2^{\Phi}=da_1^{\Phi}\,\qquad f_3^{\Phi}=\left(\frac{2}{k}\right)^{\frac{2}{3}}\star da_1^{\Phi}\\
&& f_1=da-2a_1^J-kA\,,\qquad f_4=\frac{8}{k^2}(\star da-2\star a_1^J-k\star A)\,.
\end{eqnarray}

The 5d action for the gauge fields is \cite{Cassani:2010na}

\begin{eqnarray}
\label{S}
S_{\text{5d}}&=&\frac{1}{2\kappa_5^2}\int_{M_5}\Bigg( -\left(\frac{2}{k}\right)^{\frac{2}{3}}|da_1^J|^2-\left(\frac{2}{k}\right)^{\frac{2}{3}}|da_1^{\Phi}|^2-\frac{1}{2}\left(\frac{k}{2}\right)^{\frac{4}{3}}|dA|^2-2\left(\frac{2}{k}\right)^2|da-2a_1^J-kA|^2+\nonumber \\ && +  A\wedge da_1^J\wedge da_1^J-A\wedge da_1^{\Phi}\wedge da_1^{\Phi}\Bigg)\,,
\end{eqnarray}
where

\begin{align*}
\frac{1}{2\kappa_5^2}= \frac{\text{Vol}(T^{1,1})}{2\kappa_{10}^2}=\frac{27}{32 \pi^2 k^2} N^2 \,.
\end{align*}
The equations of motion coming from eq. \eqref{S} are

\begin{align}
\label{EOMsInitial1}
&d\star da_1^J-2\left(\frac{k}{2}\right)^{\frac{2}{3}}f_4-\left(\frac{k}{2}\right)^{\frac{2}{3}}dA\wedge da_1^J=0\,,\\
&d\star dA+\left(\frac{2}{k}\right)^{\frac{4}{3}}\hspace{-0.1cm}\Big( da_1^{\Phi}\wedge da_1^{\Phi}-da_1^J\wedge da_1^J-2kf_4\Big)=0\,,\\
&d\star da_1^{\Phi}+\left(\frac{k}{2}\right)^{\frac{2}{3}}dA\wedge da_1^{\Phi}=0\,.
\label{EOMsInitial3}
\end{align}

In particular, the combination proportional to $k\,A+2\,a_1^J$ is massive due to the Stuckelberg term coming from the kinetic terms of the scalar $a$. Hence it will not give rise to a continuous symmetry, but rather to discrete 1- and 3-form symmetries, which we will ignore in the following. The R-symmetry current, on the contrary, is dual to the orthogonal combination \cite{Berenstein:2002ke}, while $a_1^{\Phi}$ does not mix with the other gauge fields and it is dual to the baryon symmetry current. In order to find the appropriate normalizations, note that $A$ must compensate for isometry transformations $\phi\rightarrow \phi+\lambda$ as 

\begin{equation}
A \rightarrow A+ \frac{2}{3} d\lambda, \qquad \oint_{M_1} d\lambda \in 2\pi\mathbb{Z}\,.
\end{equation}
This suggests that the correctly normalized $U(1)$ gauge field is $\frac{3}{2} A$, so that a convenient basis diagonalizing the kinetic terms is

\begin{equation}
\label{dictionary}
a_1^{\Phi}=\frac{2k}{3}\,A_B\,,\qquad A=\frac{2}{3}A_R+A_M\,,\qquad a_1^J=-\frac{k}{3}\, A_R+\frac{k}{4}A_M\,,
\end{equation}
where $A_R$ is the R-symmetry gauge field, $A_B$ the baryon symmetry gauge field and $A_M$ is the massive gauge field. In terms of these fields, the equations of motion in \eqref{EOMsInitial1}-\eqref{EOMsInitial3} become

\begin{align*}
&d\star dA_R+\frac{1}{2}\left(\frac{2}{k}\right)^{\frac{4}{3}}\,\Big(\frac{4k^2}{9} dA_B\wedge dA_B+\frac{3 k^2}{16} dA_M \wedge dA_M -\frac{k^2}{3} dA_R \wedge dA_R\Big)=0\, \\
&d\star dA_B+ \left(\frac{k}{2}\right)^{\frac{2}{3}} \left(\frac{2}{3} dA_R+ dA_M\right)\wedge dA_B=0\,,
\end{align*}
where we have used the first equation in \eqref{EOMsInitial1}-\eqref{EOMsInitial3} to solve for $f_4$.

In turn, the action written in terms of the $A_R,\,A_B$ fields is (we set to zero the massive gauge field $A_M$)

\begin{align}\label{symTh}
&S_{\text{SymTh}}= \frac{1}{2\pi} \int_{M_5}\Bigg(-\frac{1}{2 g_R^2}|dA_R|^2 - \frac{1}{2g_B^2}|dA_B|^2+\frac{\kappa_R}{12\pi}A_R\wedge dA_R\wedge dA_R  + \frac{\kappa_B}{4\pi} A_R\wedge dA_B\wedge dA_B \Bigg)\,,
\end{align}
with couplings

\begin{equation}
\frac{1}{g_R^2}= \frac{9N^2}{16\pi}\left(\frac{2}{k}\right)^{\frac{2}{3}}\,, \qquad  \frac{1}{g_B^2}= \frac{3N^2}{2\pi}\left(\frac{2}{k}\right)^{\frac{2}{3}}\,.
\end{equation}

From the CS level $\kappa_R, \, \kappa_B$ for $A_R$ and $A_B$ we can immediately holographically read off the 't Hooft anomalies of the Klebanov-Witten theory following \cite{Benvenuti:2006xg, Benini:2015bwz}

\begin{equation}
\kappa_R=\mathcal{A}_{RRR}=\frac{3}{2}N^2\,, \qquad \kappa_{B}= \mathcal{A}_{RBB}=-2N^2\,.
\end{equation}
Reassuringly, these match, at first order in the large $N$ expansion, the field theory expectation in \eqref{anomalies}. Notice that the reason why $k_{RRR}$ is not integer is that we assigned to the bifundamentals $A_i, \, B_i$ a minimal charge of $\frac{1}{2}$ under the R-symmetry of the Klebanov-Witten theory. However, when discussing the topological operators of the SymTh, it will be convenient to rescale the minimal charge by substituting $A_R\rightarrow 2A_R$, for the holonomies of $A_R$ to have $2\pi \mathbb{Z}$ periodicity rather than $4\pi\mathbb{Z}$.

\subsection{Charged and topological operators}\label{operators}

We  now study the charged and topological operators of the SymTh in eq. \eqref{symTh}. The action is gauge invariant (up to total derivatives) under the following gauge transformations

\begin{equation}
A_B \rightarrow A_B+ d\Lambda_B^{(0)}, \qquad A_R\rightarrow A_R+d\Lambda_R^{(0)}; \qquad \oint_{M_1} d\Lambda_B^{(0)}\in 2\pi \mathbb{Z}\,,\qquad  \oint_{M_1} d\Lambda_C^{(0)}\in 2\pi \mathbb{Z}\,.
\end{equation}

The equations of motion are

\begin{align}\label{EOMssymTh}
d\star dA_B-\frac{\kappa_Bg_B^2}{2\pi}dA_R\wedge dA_B=0, \qquad d\star dA_R -\frac{\kappa_R g_R^2}{4\pi} dA_R\wedge dA_R- \frac{\kappa_B g_R^2}{4\pi} dA_B\wedge dA_B=0\,,
\end{align}
so the following operators

\begin{align*}
&U^{(3)}_B(M_3)=e^{i\alpha_B\oint_{M_3} (g_B^{-2}\star dA_B -\frac{\kappa_B}{2\pi} A_R\wedge dA_B)},\\ &U_R^{(3)}(M_3')=e^{i\alpha_R\oint_{M_3'} (g_R^{-2}\star dA_R -\frac{\kappa_R}{4\pi} A_R\wedge dA_R-\frac{\kappa_B}{4\pi} A_B\wedge dA_B)}\,, 
\end{align*}
are topological and gauge invariant under $A_R, \, A_B$ transformations provided that

\begin{equation}
\alpha_B\in \frac{\mathbb{Z}}{\kappa_B}, \qquad  \alpha_R\in \frac{\mathbb{Z}}{\text{gcd}(\kappa_R, \kappa_B)}\,,
\end{equation}
due to the additional Chern-Simons terms.  So, in the bulk, the theory enjoys an invertible $\mathbb{Z}_{\kappa_B}\times \mathbb{Z}_{\text{gcd}(\kappa_R, \kappa_B)}$ global symmetry, under which the Wilson lines 

\begin{equation}
 W_B^{(1)}(M_1)=e^{i n_B\oint_{M_1} A_B}, \qquad W_R^{(1)}(M_1')= e^{i n_R \oint_{M_1'} A_R}\,,
\end{equation} 
are charged. 

This symmetry can be enlarged by attaching a TQFT to make the symmetry operators gauge invariant for a more general set of $\{\alpha_B, \alpha_R\}$. Examples of these extensions were discussed in \cite{Choi:2022jqy, Cordova:2022ieu, Karasik:2022kkq, GarciaEtxebarria:2022jky, Roumpedakis:2022aik, Arbalestrier:2024oqg} for theories suffering ABJ anomalies. In the following, we adopt the strategy of \cite{GarciaEtxebarria:2022jky} and we introduce a set of worldvolume scalars $\rho_R$ for $U_B^{(3)}$ and $\rho_R',\, \rho_B$ for $U_R^{(3)}$ transforming under the gauge transformations as

\begin{equation}
\rho_B\rightarrow \rho_B- \Lambda_B^{(0)}, \qquad \rho_R\rightarrow \rho_R- \Lambda_R^{(0)},\qquad \rho_R'\rightarrow  \rho_R'- \Lambda_R^{(0)}\,,
\end{equation}
such that the dressed operators

\begin{eqnarray}
&&U_B^{(3)}(M_3)= e^{i \alpha_B \oint_{M_3} g_B^{-2}\star dA_B}\int \mathcal{D}\rho_R\, e^{-i\alpha_B \frac{\kappa_B}{4\pi}\oint_{M_3}(A_R+d\rho_R)\wedge dA_B}, \\ \nonumber
&&U_R^{(3)}(M_3')= e^{i\alpha_R \oint_{M_3'} g_R^{-2}\star dA_R }\int \mathcal{D}\rho'_R \mathcal{D}\rho_B e^{-i\alpha_R\oint_{M_3'} \left(\frac{\kappa_B}{4\pi} (A_B+d\rho_B)\wedge dA_B+\frac{\kappa_R}{4\pi} (A_R+d\rho'_R)\wedge dA_R\right)}\,, \\ &&
\end{eqnarray}
are gauge invariant for any $\alpha_B, \, \alpha_R$.

The drawback of this procedure is that, although the action of these operators on $W_B^{(1)}, \, W_R^{(1)}$ is fully invertible, they act as projectors on configurations with non-trivial fluxes of $A_B$ and $A_B, \, A_R$ respectively. In other words, correlation functions between these operators and any 't Hooft lines for $A_B, \, A_R$ are trivial.

The Bianchi identities $d^2A_B=d^2A_R=0$ can be regarded as conservation equations of 2-form currents. The corresponding topological operators 

\begin{align*}
V_B^{(2)}(M_2)=e^{i \beta_B \oint_{M_2} dA_B}, \qquad V_R^{(2)}(M_2')= e^{i\beta_R \oint_{M_2'} dA_R}\,,
\end{align*}
are gauge invariant and topological for any\footnote{The parameters are compact since the flux is quantized, so $\beta_B \sim \beta_B+2\pi$ and similarly for $\beta_R$.} $\beta_B, \, \beta_R \in U(1)$.  These operators measure the charge of the 't Hooft lines, which can be described as holonomies of the field Hodge dual to $A_B, \, A_R$. These fields $C_B, \, C_R$ can be constructed from solving the equations of motion as

\begin{equation}\label{dualbaryon}
g_B^{-2}\star dA_B - \frac{\kappa_B}{2\pi} A_R\wedge dA_B= dC_B, \qquad g_R^{-2} \star dA_R-\frac{\kappa_R}{4\pi} A_R\wedge dA_R-\frac{\kappa_B}{4\pi} A_B\wedge dA_B= dC_R\,.
\end{equation}
Since $dA_B, \, dA_R$ are gauge invariant, we see that the fields $C_B, \, C_R$ are 2-form gauge fields transforming as

\begin{equation}
C_B\rightarrow C_B+ d\Lambda_B^{(1)}- \frac{\kappa_B}{2\pi} \Lambda_R^{(0)} dA_B, \qquad  C_R\rightarrow C_R+ d\Lambda_R^{(1)} - \frac{\kappa_R}{4\pi} \Lambda_R^{(0)} dA_R- \frac{\kappa_B}{4\pi} \Lambda_B^{(0)} dA_B\,,
\end{equation}
under the $U(1)_R$ and $U(1)_B$ gauge transformations. These anomalous transformations imply that the naive 't Hooft lines

\begin{equation}
T_B^{\text{naive}}(N_2)=e^{i m_B \oint_{N_2} C_B}, \qquad T_R^{\text{naive}}(N_2')=e^{i m_R \oint_{N_2'} C_R}\,,
\end{equation}
are not gauge invariant. To cure this problem we can dress these lines by attaching a 3d TQFT as in \cite{Kaidi:2021xfk, Bergman:2024its}. Just like above, we choose instead to attach a TQFT as

\begin{align*}
&T_B^{(2)}(N_2)=\int \mathcal{D}\rho_R\,e^{i m_B \oint_{N_2} \left( C_B-\frac{\kappa_B}{2\pi} \rho_R\,dA_B\right)}, \\ &T_R^{(2)}(N_2')=\int \mathcal{D}\rho_R' \mathcal{D}\rho_B\,e^{i m_R \oint_{N_2'} \left( C_R-\frac{\kappa_B}{4\pi} \rho_B \, dA_B-\frac{\kappa_R}{4\pi} \rho_R' \,dA_R\right)}\,,
\end{align*}
where $\rho_B, \rho_R, \rho_R'$ represent worldvolume scalars transforming as 
\begin{equation}
 \rho_B\rightarrow \rho_B- \Lambda_B^{(0)}, \qquad \rho_R \rightarrow \rho_R-\Lambda_R^{(0)},\qquad  \rho_R' \rightarrow \rho_R'-\Lambda_R^{(0)}\,,
\end{equation}
under the gauge transformations. 

Before analyzing the linkings among the operators and the possible set of boundary conditions, let us mention that 5d Chern-Simons theories enjoy also instantonic symmetries \cite{Seiberg:1996bd, Arbalestrier:2025poq}. The corresponding current is topologically conserved and the corresponding charges are disorder operators. Since we are focusing mainly on the role of the R- and baryonic symmetries, we will not discuss these symmetries or their possible brane realization in the following, leaving their study to future work. 

\subsubsection{Linking and boundary conditions}\label{Linkingandbcs}

Having identified the charged and symmetry operators, we can now study their linking and possible constraints on the boundary conditions for the SymTh. 

Consider the insertion of the Wilson lines $W_B^{(1)}$ and $W_R^{(1)}$ in the path integral. The equations of motion are sourced as

\begin{align*}
&\frac{1}{g_B^2}d\star dA_B-\frac{\kappa_B}{2\pi}dA_R\wedge dA_B=-n_B\, \delta(M_1), \\
&\frac{1}{g_R^2}d\star dA_R-\frac{\kappa_R}{4\pi} dA_R\wedge dA_R- \frac{\kappa_B}{4\pi} dA_B \wedge dA_B=-n_R\, \delta(M_1')\,,
\end{align*}
indicating a non-trivial linking if $W^{(1)}_B(M_1)$ (resp. $W^{(1)}_R(M_1')$) is surrounded by $U^{(3)}_B(M_3)$ (resp. $U^{(3)}(M_3')$). In other words, there are two 1-form symmetries under which the Wilson lines are charged. 

Similarly, the 't Hooft lines source the Bianchi identities, leading to a non-trivial linking between $V_B^{(2)}(M_2)$ and $T_B^{(2)}(N_2)$ and $V_R^{(2)}(M_2')$ and $T_R^{(2)}(N_2')$. They are then charged under two 2-form symmetries generated by $V_B^{(2)}(M_2)$ and $V_R^{(2)}(M_2')$.

The kinetic and Chern-Simons terms introduce constraints on the possible boundary conditions for the supergravity fields. 
Kinetic terms require opposite boundary conditions for the gauge fields $A_B, \, A_R$, and their duals: if, say, $A_B$ is \textbf{D}irichlet, the dual field $C_B$ is \textbf{N}eumann and viceversa.\footnote{Note that here we are consider the SymTh in flat space. In $AdS$, the standard $AdS/CFT$ dictionary requires also normalizability of the fluctuations, which for gauge fields in $AdS_5$ means to impose Dirichlet boundary conditions \cite{Marolf:2006nd, Das:2022auy}. Yet, in the spirit of the SymTh we are only interested on symmetries and so we will adopt the approach of \cite{DeWolfe:2020uzb, Damia:2022bcd}.} So, in the absence of Chern-Simons terms, we are left with four possible boundary conditions for $(A_B, A_R)$: $(\textbf{D}, \textbf{D}), \, (\textbf{N},\textbf{D}), \, (\textbf{D}, \textbf{N}), \, (\textbf{N},\textbf{N})$. When the Chern-Simons terms are considered, the triple self-anomaly forbids Neumann boundary conditions for $A_R$  \cite{Damia:2022bcd}. This is the SymTh realization of the self 't Hooft anomaly of the R-symmetry in the Klebanov-Witten theory, since imposing Neumann boundary conditions for $A_R$ would have meant gauging the R-symmetry \cite{Apruzzi:2024htg}. We are then left with two possible boundary conditions:

\begin{enumerate}
\item $(\textbf{D},\textbf{D})$: Dirichlet boundary conditions for the gauge fields imposes $dA_B=dA_R=0$ at the boundary. Both $W_B^{(1)}(M_1)$ and $W_R^{(1)}(M_1')$ are then allowed to end at the boundary, while the dual 't Hooft lines ending at the boundary describe non-genuine operators. In turn, the symmetry operators $V_B^{(2)}(M_2)$ and $V_R^{(2)}(M_2')$ are trivialized, while the topological operators $U_B^{(3)}(M_3)$ and $U_R^{(3)}(M_3')$ are non-trivial at the boundary and measure the charge of  $W_B^{(1)}(M_1)$ and $W_R^{(1)}(M_1')$. Moreover, since $A_B, \, A_R$ enjoy Dirichlet boundary conditions, the operators become invertible

\begin{align*}
&U_B^{(3)}(M_3)\sim  e^{i \alpha_B \oint_{M_3} g_B^{-2}\star dA_B}, \\
&U_R^{(3)}(M_3')\sim e^{i\alpha_R \oint_{M_3'} g_R^{-2}\star dA_R }\,.
\end{align*}

This choice of boundary condition corresponds to the original KW theory, where there are two invertible 0-form symmetries: the baryonic symmetry and the R-symmetry. The corresponding operators are $U^{(3)}(M_3)$ and $U^{(3)}(M_3')$, while the genuine charged ones are $W_B^{(1)}(M_1)$ and $W^{(1)}(M_1')$. The R-symmetry has a self 't Hooft anomaly and a mixed 't Hooft anomaly with the baryonic symmetry. Both anomalies are described in the bulk by the Chern-Simons terms, describing the anomaly theory.

\item $(\textbf{N},\textbf{D})$: Dirichlet boundary conditions for $A_R$ still imply $dA_R=0$, while Neumann boundary conditions for $A_B$ imply that $C_B$ is Dirichlet. The charged operators $W_R^{(1)}(M_1')$ and $T_B^{(2)}(N_2)$ ending at the boundary describe genuine operators of corresponding QFT, while $W_B^{(1)}(M_1)$ and $T_R^{(2)}(N_2')$ realize non-genuine operators. Moreover, $V_R^{(2)}(M_2')$ still trivialize at the boundary, while $V_B^{(2)}(M_2)$ is non-trivial and measures the charge of $T_B^{(2)}(N_2)$.

The topological operator $U_B^{(3)}(M_3)$ trivializes at the boundary due to the Dirichlet boundary conditions for the dual field. On the contrary, $U_R^{(3)}(M_3')$ does not trivialize and it is non-invertible

\begin{equation}
U_R^{(3)}(M_3')\sim  e^{i\alpha_R \oint_{M_3'} g_R^{-2}\star dA_R }\int \mathcal{D}\rho_B\,  e^{-i\alpha_R \frac{\kappa_B}{4\pi}\oint_{M_3'} (A_B+d\rho_B) dA_B}\,.
\end{equation}

This choice of boundary conditions corresponds to the gauging of the baryonic symmetry in the Klebanov-Witten theory. Indeed, as discussed above, upon gauging the baryonic symmetry we obtain a new 1-form magnetic symmetry. Moreover, the R-symmetry now suffers an ABJ anomaly and becomes non-invertible. 

\end{enumerate}

\section{Branes and operators}\label{Section4}

We now describe the string theory realization of the charged and symmetry operators of the SymTh as branes wrapped on $T^{1,1}$. We collect in Appendix \ref{AppendixConifold} some details of its geometry, in particular pertaining 2- and 3-cycles (to which we will refere as $S^2$ and $S^3$). We will first focus on the baryonic symmetry and then we will discuss the R-symmetry case. To compare with the Klebanov-Witten theory, we are expressing the worldvolume action in terms of the field $A_B$ and the $4\pi$ periodic R-symmetry field in \cref{symTh}. 

\subsection{Baryon symmetry}\label{baryonsymmetry}

As discussed in \cite{Bergman:2024aly}, the holographic realization of the baryonic symmetry operators are non-BPS D4 branes wrapped on the 2-cycle of the $T^{1,1}$. A representative of such cycle can be taken at $\theta_2=\pi-\theta_1=\theta$, $\phi_1=\phi_2=\phi$ \cite{Arean:2004mm}. The pullback of $J$ then vanishes, so the part of $F_5$ which can contribute to the D4 brane WZ term is

\begin{equation}
F_5\supset \frac{4}{3}\left(\frac{k}{2}\right)^{\frac{1}{3}}\left(\star dA_B+\frac{2}{3}\left(\frac{k}{2}\right)^{\frac{2}{3}} dA_B\wedge A_R\right)\wedge \Phi\,.
\end{equation}

The worldvolume action of a D4 brane in the tachyon vacuum becomes 

\begin{equation}
\label{SD4}
S_{\text{D4}}^{\text{WZ}}= \alpha T_{\text{D4}} \int_{M_3\times S^2} F_5=\alpha \frac{16\pi T_{\text{D4}}}{9}\left(\frac{k}{2}\right)^{\frac{1}{3}}  \, \int_{M_3} \left(\star dA_B+\frac{2}{3}\left(\frac{k}{2}\right)^{\frac{2}{3}} dA_B\wedge A_R\right)\,.
\end{equation}
Note that, in addition to the first term $\star dA_B$ considered in \cite{Bergman:2024aly}, there is an extra term quadratic in the gauge fields coupling to $A_R$. 

While it is clear that the operator defined  by the D4 is topological by virtue of the equations of motion \eqref{EOMsInitial1}-\eqref{EOMsInitial3}, \eqref{SD4} shows that it is not gauge invariant under the gauge transformation of $A_R\rightarrow A_R+d\lambda$ for arbitrary $\alpha$. Since the $A$ gauge transformations should be accompanied by diffeomorphism shifting $\phi\rightarrow \phi +\lambda$, given our D4 sits at a fixed value of $\phi$, the isometry is broken explicitly. This suggests that, in order to recover gauge-invariance of the worldvolume action, we need to consider a generic embedding where the transverse coordinate to the brane along the fiber is dynamical. As a consequence, there is an extra term from the pull-back of $F_5$. All in all, the full action on the D4 is

\begin{equation}
\label{SD4FULL}
S_{\text{D4}}^{\text{WZ}}= \alpha\, \int_{M_3} \left(g_B^{-2}\star dA_B-\frac{\kappa_B}{2\pi}(A_R+d\rho_R)\wedge dA_B\right)\,,
\end{equation}
where we are choosing a convenient normalization for the transverse scalar $\rho_R$ such that $\rho_R \rightarrow \rho_R-\lambda$ and we reabsorbed an overall numerical coefficients in $\alpha$. 

 Let us stress that $\rho_R$ is a worldvolume field, so the worldvolume theory of the D4 will include a path integral over $\rho_R$ on the defect. This explicitly realizes the chosen dressing of symmetry operators in \cref{operators}.

The operator charged under this symmetry is a D3 brane wrapping the homological non-trivial $S^3$ of the conifold. We can take such $S^3$  to be spanned by $\theta_1,\phi_1,\psi$ at  $\theta_2=\phi_2=0$. From \eqref{F5} the relevant part of $F_5$ is

\begin{equation*}
F_5\supset \frac{2k}{3}dA_B\wedge \Phi\wedge \eta-\frac{k}{3}dA_R\wedge J\wedge \eta=-d\left(\frac{2k}{3}\frac{\sin\theta_1}{18}\left(A_B-\frac{1}{2} A_R\right)\wedge d\psi\wedge d\theta_1\wedge d\phi_1 \right)\equiv dC_4\,.
\end{equation*}
Thus, the WZ term of the D3 is

\begin{equation}
S_{\text{D3}}^{\text{WZ}}=\frac{16\pi^2 k T_{\text{D3}}}{27}\int_{M_1} \left( A_B-\frac{1}{2}\,A_R\right)\,,
\end{equation}
being
\begin{equation}
\int_{S^3} \Phi \wedge \eta= -\frac{8\pi^2}{9}\,.
\end{equation}
Since $T_{\text{D3}}=\frac{1}{(2\pi)^3}$, we see that

\begin{equation}
S_{\text{D3}}^{\text{WZ}}=N\,\int_{M_1} A_B-\frac{N}{2}\,\int_{M_1} A_R\,.
\end{equation}
Indeed, the R-charge is the expected one for the dibaryon, since $\mathcal{B}={\rm det}A_1$, so $\Delta[\mathcal{B}]=N\times \frac{1}{2}$ and $R[A_1]=\frac{1}{2}$. 

\subsubsection{Dual magnetic symmetry}

The dual magnetic 1-form symmetry obtained from gauging the baryonic symmetry is also implemented by a D4 brane, this time wrapping the $S^3$ of the $T^{1,1}$. The relevant terms of the $F_5$ contributing when pullbacked on the D4 are still 

\begin{equation}
F_5\supset  \frac{2k}{3} dA_B\wedge \Phi\wedge \eta-\frac{k}{3}A_R\wedge J\wedge \eta\,,
\end{equation}
and the calculation proceeds as in the D4 case, leading to (after absorbing an overall numerical coefficient in the parameter $\alpha$)

\begin{equation}
S_{\text{D4}}^{\text{WZ}}=\alpha\int_{M_2}\left(dA_B-\frac{1}{2}dA_R\right)\,.
\end{equation} 
We then see that the D4 brane realizes a symmetry operator measuring both the charge of $T_B^{(2)}(N_2)$ and $T_R^{(2)}(N_2')$. This is to be expected given that the D3 brane wrapping $S^3$ is charged under both the baryonic and the R-symmetry. 

We can similarly obtain the 't Hooft line $T_B^{(2)}(N_2)$ as a D3 brane wrapping the $S^2$ of the $T^{1,1}$. The pullback of the $F_5$ field strength reads

\begin{align*}
P[F_5]=\frac{2k}{3}\left(\frac{2}{k}\right)^{\frac{2}{3}} \star dA_B\wedge \Phi+ \frac{4k}{9}\left(A_R+d\rho_R \right)dA_B \wedge \Phi\,.
\end{align*}

This is closed, since from eq. \eqref{dualbaryon}

\begin{equation}
\star dA_B=\frac{2\pi}{3N^2} \left(\frac{k}{2}\right)^{\frac{2}{3}} dC_B-\frac{2}{3}\left(\frac{k}{2}\right)^{\frac{2}{3}}A_R dA_B\,.
\end{equation}
Then $P[C_4]$ reads

\begin{equation}
P[C_4]= \frac{4\pi k}{9N^2}  C_B\wedge \Phi+ \frac{4k}{9} \rho_R dA_B\wedge \Phi\,,
\end{equation}
leading to the brane action

\begin{equation}
S_{\text{D3}}^{\text{WZ}}=\frac{16\pi^2k}{27} T_{\text{D3}} \int_{N_2} \left(\frac{1}{N^2}C_B+\frac{1}{\pi}\rho_R dA_B \right)= \frac{1}{N}\int_{N_2} \left(C_B-\frac{\kappa_B}{2\pi}\rho_R dA_B \right)\,.
\end{equation}
This realizes the 't Hooft line as expected.\footnote{Notice that these lines come in multiples of $\frac{1}{N}$. This is expected, since the minimal Wilson line for the baryonic field has charge $N$ and the two set of charged operators are dual. }

\subsection{R-symmetry}

Let us finally consider the R-symmetry operators. Following the discussion of section \ref{Section1}, we expect a non-BPS KK monopole with Taub-NUT direction $\phi$ to implement the R-symmetry. Before delving into the details of this, we need to complete the action of the non-BPS KK monopole constructed in \cref{Section1} to the full sector sensitive to the closed string fields of interest for us (namely, metric and RR 4-form potential). The BPS KK$_B$ monopole contains a worldvolume 2-form potential $\omega^{(2)}$ which appears in the action through  \cite{Eyras:1998hn} 

\begin{equation}
S_{\text{KK}_B}^{\text{WZ}}= 4\pi^2 T_{\text{KK}_B}\int_{M_7} d\omega^{(2)} \wedge \mathcal{K}_3\,, \qquad \mathcal{K}_3=d\omega^{(2)}+\frac{1}{2\pi} \imath_{\kappa}C_4\,,
\end{equation}
so that the BPS monopole action is invariant under the gauge transformations
\begin{align*}
&\imath_{\kappa}\mathcal{N}\rightarrow \imath_{\kappa}\mathcal{N}+d(\imath_{\kappa}\Sigma_6)-\frac{1}{2} d(\imath_{\kappa}\Lambda_3)\wedge \imath_{\kappa}C_4\,,\\
&C_4\rightarrow C_4+d\Lambda_3\,,\\
&d\omega^{(2)}\rightarrow d\omega^{(2)}-\frac{1}{2\pi}d\left(\imath_{\kappa}\Lambda_3\right)\,.
\end{align*}
By consistency, the full non-BPS KK monopole action before the tachyon decay must then be

\begin{equation}
S_{\overline{\text{KK}}_B}^{\text{WZ}}=T_{\overline{\text{KK}}_B}\int_{M_7} Z(T)\,\left[d(\imath_{\kappa}\mathcal{N})+\frac{1}{2}\,(2\pi)\,d\omega^{(2)}\wedge \imath_{\kappa}F_5\right]\,.
\end{equation}
As described above, strictly speaking, $d(\imath_{\kappa}\mathcal{N})$ picks extra contributions from RR potentials as follows

\begin{equation}
d(\imath_{\kappa}\mathcal{N})=e^{-2\tilde{\Phi}}\star d\tilde{B}_2+\frac{1}{2} \imath_{\kappa}C_4\wedge \imath_{\kappa}F_5\,.
\end{equation}

We are now ready to study non-BPS KK monopoles in our background. The metric of Appendix \ref{AppendixConifold} with $A$ fluctuations reads

\begin{equation*}
ds^2=\left(\frac{2}{k}\right)^{\frac{5}{6}}\,ds_5^2+\left(\frac{k}{2}\right)^{\frac{1}{2}}\,\left[ \frac{4}{9}\left(d\phi+\frac{1}{2}\cos\theta_id\phi_i-\frac{3A}{2}\right)^2+\frac{1}{6}(d\theta_1^2+\sin^2\theta_1d\phi_1^2)+\frac{1}{6}(d\theta_2^2+\sin^2\theta_2d\phi_2^2)\right]\,.
\end{equation*}

To determine $\imath_{\kappa}\mathcal{N}$ in terms of IIB fields, we T-dualize along $\phi$

\begin{align*}
&d\tilde{s}^2= \left(\frac{2}{k}\right)^{\frac{5}{6}}\,ds_5^2+\frac{9}{4}\left(\frac{2}{k}\right)^{\frac{1}{2}} d\phi^2 +\left(\frac{k}{2}\right)^{\frac{1}{2}}\,\Big[\frac{1}{6}(d\theta_1^2+\sin^2\theta_1d\phi_1^2)+\frac{1}{6}(d\theta_2^2+\sin^2\theta_2d\phi_2^2)\Big]\,,\\
&\tilde{B}_2=\left(\frac{3A}{2}-\xi\right)\wedge d\phi, \qquad e^{-2\tilde{\Phi}}= \frac{4}{9} \left(\frac{k}{2}\right)^{\frac{1}{2}}, \qquad \tilde{C}_3=-\imath_{\kappa}C_4\,,
\end{align*}
with $\xi= \frac{1}{2}\cos\theta_i d\phi_i$. As a consequence (dropping the field-independent contribution), the Hodge dual of $\tilde{H}_3$ reads
\begin{equation}
\star d\tilde{B}_2=\frac{1}{2}\left(\frac{k}{2}\right)^{\frac{5}{6}}\tilde{\star}_{AdS_5} dA\wedge \Phi \wedge \Phi\,.
\end{equation}

In turn

\begin{equation}
\imath_{\kappa} F_5=-\frac{2}{3} da_1^\Phi\wedge \Phi-\frac{2}{3} da_1^J\wedge J-\frac{2k}{3}\,J\wedge J-\frac{2}{3}f_4\,.
\end{equation}
It then follows

  \begin{equation}
 \imath_{\kappa}C_4\wedge \imath_{\kappa}F_5=\frac{4}{9}\,\Big(a_1^{\Phi}\wedge da_1^{\Phi}-a_1^J\wedge da_1^J +k\,A\wedge da_1^J-2^{\frac{2}{3}}\, k^{\frac{1}{3}}\,\star da_1^J \Big) \wedge \Phi\wedge \Phi\,,
 \end{equation}
 where we have used the SUGRA equations of motion to solve for $f_4$.

Finally, we reduce the 2-form as\footnote{The other components of $\omega^{(2)}$ do not contribute to the worldvolume theory of the non-BPS KK monopole. }

\begin{equation}
2\pi\,d\omega^{(2)}=-\frac{2}{3}\left(\frac{2k}{3}\,d\rho_B\wedge \Phi-k\,d\rho_R\wedge J\right)\,,
\end{equation}
where we have fixed the normalization of the scalars $\rho_{B,R}$ for latter convenience. Putting together all ingredients and using the dictionary in \cref{dictionary}, the action for the non-BPS KK monopole is

\begin{equation}
S_{\overline{\text{KK}}_B}^{\text{WZ}}=\alpha\,\int_{M_3'} \left(g_R^{-2}\star dA_R-\frac{\kappa_B}{4\pi}(A_B+d\rho_B)\wedge dA_B-\frac{\kappa_R}{4\pi}(A_R+d\rho_R)\wedge dA_R \right)\,.
\end{equation}

The operator charged under the R-symmetry is dual to a gravitational wave propagating in the isometry direction. Focusing on the WZ term in the minimal case $m=1$

\begin{equation}
S_{\text{GW}}^{\text{WZ}}=\frac{3}{2} \int_{M_1'} A= \int_{M_1'} A_R\,,
\end{equation}
the gravitational wave reduces to a Wilson line of $A_R$. In particular, the charge under the R-symmetry is integer, contrary to the baryons whose R-charge is in general fractional. This reproduces the fact that mesons contain an even number of elementary fields of half-integer R-charge and hence have an integer R-charge (as opposed to the the R-charge of baryons which is $\frac{N}{2}$ and thus generically half-integer).

\subsubsection{Dual objects}

Finally, let us consider the objects dual to the R-symmetry charged and topological operators. These are respectively BPS KK monopoles and NS2$_R$ branes. 

Let us first consider a KK monopole wrapping the base $S^2\times S^2$ of $T^{1,1}$ with isometry direction $\phi$. This object couples to $\imath_{\kappa}\mathcal{N}$, which can be read off from

\begin{align*}
&d\left(\imath_{\kappa}\mathcal{N}\right)=\frac{4}{9}\left(\frac{k}{2}\right)^{\frac{4}{3}}\left(\tilde{\star}_{\text{AdS}_5} dA_R+ \frac{1}{2}\left(\frac{2}{k}\right)^{\frac{4}{3}}\Big(a_1^{\Phi}\wedge da_1^{\Phi}-a_1^J\wedge da_1^J +k\,A\wedge da_1^J\Big)\right)\wedge \Phi \wedge \Phi=\\
&=\frac{4}{9} \left(\frac{k}{2}\right)^{\frac{4}{3}} g_R^2\,dC_R\wedge \Phi \wedge \Phi\,.
\end{align*}
As before, we expand the worldvolume gauge field $\omega^{(2)}$ as

\begin{equation}
2\pi d\omega^{(2)} =-\frac{2}{3}\left(\frac{2k}{3}\,d \rho_{B}\wedge \Phi-k\, d\rho_J \wedge J\right)\,,
\end{equation}
where we again choose the normalization for latter convenience. Putting all ingredients together and using \cref{dictionary}, the KK action is

\begin{equation}
S_{\text{KK}_B}^{\text{WZ}}= T_{\text{KK}}\int_{N_6} \left(\imath_{\kappa} \mathcal{N} -\frac{1}{2} 2\pi\,d\omega^{(2)} \wedge \imath_{\kappa}C_4\right)= \int_{N_2'} \left(C_R- \frac{\kappa_B}{4\pi} \rho_B dA_B - \frac{\kappa_R}{4\pi} \rho'_R dA_R\right)\,,
\end{equation}
which is precisely the R-symmetry 't Hooft line.

Finally, the object measuring the charge of the BPS KK monopole is a NS2$_R$, which action is 
\begin{equation}
S_{\text{NS2}_R}^{\text{WZ}}=\sqrt{2}\alpha \int_{M_2'} d(\imath_{\kappa}\tilde{B})= \alpha\int_{M_2'} dA_R\,,
\end{equation}
reproducing the magnetic dual operator to the non-BPS KK monopole. 

\subsection{Brane-operator matching}

Having discussed the spectrum of branes corresponding to charged and symmetry operators for the baryon and R-symmetries, we can now match them to our SymTh analysis. The precise map among the charged and topological operators and the corresponding BPS and non-BPS branes is reported in Table \ref{Table1}.

\begin{table}[h!]
\centering
\begin{tabular}{ |c|c|c|} 
\hline
Topological operator & non-BPS brane \\ 
 \hline
 $U_B^{(3)}(M_3)$ & $\text{D4}|_{S^2\times M_3}$\\
 $U_R^{(3)}(M_3')$ & $\overline{\text{KK}}_B|_{S^2\times S^2\times M_3'}$\\
 $V_B^{(2)}(M_2)$ & $\text{D4}|_{S^3\times M_2}+ \text{NS2}_R|_{M_2}$\\
 $V_R^{(2)}(M_2')$ & $\text{NS2}_R|_{M_2'}$\\
 \hline
 Charged operator & BPS brane \\ 
 \hline
 $W_B^{(1)}(M_1)$ & $\text{D3}|_{S^3\times M_1}+\left[\frac{N}{2}\right]\text{GW}|_{M_1}$\\
 $W_R^{(1)}(M_1')$ & $\text{GW}|_{M_1'}$\\
 $T_B^{(2)}(N_2)$ & $\text{D3}|_{S^2\times N_2}$\\
 $T_R^{(2)}(N_2')$ & $\text{KK}_B|_{S^2\times S^2 \times N_2'}$\\
 \hline
\end{tabular}
\caption{Match between branes and SymTh operators.}\label{Table1}
\end{table}

Notice that if $N$ is even, we can construct objects that are neutral under the R-symmetry by combining a D3 with $\left[\frac{N}{2}\right]$ GW. This matches the expectations from the Klebanov-Witten theory: the baryon has an R-symmetry charge $\frac{N}{2}$, while mesons have always an integer R-charge. We can then combine $\left[\frac{N}{2}\right]$ mesons with a single baryon, obtaining an operator with $\frac{N}{2}-\left[\frac{N}{2}\right]$ under the R-symmetry, such that if $N$ is even the charge is zero. We remark that, compatibly with the Klebanov-Witten spectrum, the minimal D3 brane has charge $N$ under the 1-form symmetry.

\section{Conclusions}\label{Conclusions}

In this paper we have proposed a brane realization of symmetry operators corresponding to holographic QFTs' symmetries associated to the geometry of the dual gravitational background. The proposed brane is a non-BPS Kaluza-Klein monopole, whose existence is required by dualities and whose worldvolume action can be constructed by consistency with the well-studied action for BPS KK monopoles. As we have shown, non-BPS KK monopoles naturally link with gravitational waves propagating along the Taub-NUT direction of the non-BPS monopole.

Using this proposal we have been able to describe in full detail the baryon+R-symmetry sector of the Klebanov-Witten theory, completing the picture in \cite{Bergman:2024aly}. As proposed in that reference, the topological operators for the baryon symmetry are non-BPS D4 branes on the $S^2$ of the $T^{1,1}$ which link with BPS D3 branes on the $S^3$ of the $T^{1,1}$.  Exchanging them --that is, the D4 on the $S^3$ linking with the D3 on the $S^2$-- corresponds to the dual magnetic symmetry topological and charged operators. In turn, the topological operators for the R-symmetry are non-BPS KK monopoles with Taub-NUT direction along the Reeb vector of the $T^{1,1}$ and link with GW propagating along the same direction --as well as baryon D3 branes. Using the full Kaluza-Klein reduction of Type IIB on the $T^{1,1}$, it turns out that the non-BPS branes precise reproduce the expected behavior of the symmetry operators as read off from the SymTh, including the correct 't Hooft anomalies. Moreover, the R-symmetry operators naturally come with a TQFT dressing, arising from worldvolume fields on the branes, which renders them non-invertible for some specific boundary conditions.

Symmetries of holographic QFTs associated to isometries are often non-abelian. This includes both ordinary mesonic symmetries (such as the $SU(2)\times SU(2)$ global symmetry of the Klebanov-Witten theory acting on mesons) as well as R-symmetries in theories with extended supersymmetry. Due to their non-abelian structure, the description of these cases appears much more complicated. It would be extremely interesting to investigate the description of these operators, perhaps in terms of non-BPS KK monopoles, and how the non-abelian structure can be detected.

We have described the baryon+R-symmetry sector of the Klebanov-Witten theory using the SymTh arising from the reduction of Type IIB Supergravity on $T^{1,1}$. This SymTh is a   5d Chern-Simons Maxwell theory obtained from truncating to the gauge field sector of \cite{Cassani:2010na}. The motivation for this is that we are mostly interested on the near-boundary physics where our symmetry operators live. On the other hand, going close to the AdS boundary is effectively probing the deep IR of the theory. Borrowing arguments from the non-abelian case in \cite{Bonetti:2024cjk}, it is natural to expect that the SymTh goes over the SymTFT for two $U(1)$s with a set of mixed anomalies \cite{Antinucci:2024zjp, Arbalestrier:2025poq}

\begin{align*}
S_{\text{SymTFT}}= \frac{1}{2\pi}\int_{M_5} \left(f_R\wedge dA_R+f_B\wedge dA_B + \frac{\kappa_B}{4\pi}A_R\wedge dA_B\wedge dA_B + \frac{\kappa_R}{12\pi} A_R\wedge dA_R\wedge dA_R \right)\,,
\end{align*}
where the $f_R, \, f_B$ are $\mathbb{R}$ gauge fields associated with the Hodge dual of the field strengths $dA_R, \, dA_B$.  

It would be very interesting to match this SymTFT to our description of the SymTh operators in terms of String Theory branes.

\section*{Acknowledgments}

We would like to thank O. Bergman, A. Faedo, E. Garc\'ia-Valdecasas and Y. Lozano for discussions.  The authors are supported in part by the Spanish national grant MCIU-22-PID2021-123021NB-I00.

\begin{appendix}

\section{Conifold geometry}\label{AppendixConifold}

In this Appendix, we review the metric of the solution following Appendix $B$ of \cite{Cassani:2010na}. The vierbeins are

\begin{eqnarray}
e_1&=&-\sin\theta_1\,d\phi_1\,,\nonumber \\
 e_2&=&d\theta_1\,,\nonumber \\
 e_3 &=& \cos\psi\,\sin\theta_2\,d\phi_2-\sin\psi\,d\theta_2\,,\nonumber \\
 e_4&=&\sin\psi\,\sin\theta_2\,d\phi_2+\cos\psi\,d\theta_2\,,\nonumber \\
 e_5&=&d\psi+\cos\theta_1\,d\phi_1+\cos\theta_2\,d\phi_2\,,\nonumber 
 \end{eqnarray}
 where the angles have periodicities $\phi_i\in [0, 2\pi),\, \theta_i\in [0, \pi)$ and $\psi\in [0, 4\pi)$.

 The solution is an unwarped product of an $AdS_5$ and a $T^{1,1}$ with metric
 \begin{equation}
ds^2=\left(\frac{2}{k}\right)^{\frac{5}{6}}\,ds_{AdS_5}+\left(\frac{k}{2}\right)^{\frac{1}{2}} ds^2_{T^{1,1}}\,,
\end{equation}
where the metric of the unperturbed conifold  reads
\begin{equation}
ds_{T^{1,1}}^2=\frac{1}{9}(d\psi+\cos\theta_1\,d\phi_1+\cos\theta_2\,d\phi_2)^2+\frac{1}{6}(d\theta_1^2+\sin\theta_1^2d\phi_1^2)+\frac{1}{6}(d\theta_2^2+\sin\theta_2^2d\phi_2^2)\,.
\end{equation}

The base of the conifold enjoys a second and third non-trivial cohomology, which generators can be written explicitly as follows. Firstly, define the 2-forms

\begin{align*}
&J=\frac{1}{6}(e^{12}-e^{34})=\frac{1}{6}(\sin\theta_1\,d\theta_1\wedge d\phi_1+\sin\theta_2\,d\theta_2\wedge d\phi_2)\,,\\
&\Phi=\frac{1}{6}(e^{12}+e^{34})=\frac{1}{6}(\sin\theta_1\,d\theta_1\wedge d\phi_1-\sin\theta_2\,d\theta_2\wedge d\phi_2)\,,
\end{align*}
where $\Phi, \, J$ are both closed, but $J$ is exact
\begin{equation}
d\eta=2J, \,\,\,\,\,\, \eta= -\frac{1}{3}(d\psi + \cos\theta_1d\phi_1+ \cos\theta_2d\phi_2)\,.
\end{equation}

Then $\Phi$ defines the generator of the second cohomology class, while the third cohomology is generated by $\Phi \wedge \eta$. A representative of the 2-cycle can be taken at $\theta_2=\pi-\theta_1=\theta$, $\phi_1=\phi_2=\phi$ (see \textit{e.g.} \cite{Arean:2004mm}), where 

\begin{equation}
J=0\,,\qquad \int_{S^2} \Phi= \frac{4\pi}{3}\,.
\end{equation}
In turn, the baryonic 3-cycle sits at constant $(\theta_2,\,\phi_2)$ \cite{Gubser:1998fp} (see also \cite{Arean:2004mm}). It is then immediate to calculate its volume integrating the 3-form

\begin{equation}
\Phi\wedge \eta = J\wedge \eta =-\frac{\sin\theta_1}{18}\,d\psi\wedge d\theta_1\wedge d\phi_1\,,\qquad \left|\int\Phi\wedge \eta\right| = \left|\int J\wedge \eta\right|=\frac{8\pi^2}{9}\,.
\end{equation}

Finally, another useful formula is the integral of $\Phi \wedge \Phi$ over the base of $T^{1,1}$, which reads

\begin{equation}
\int_{S^2\times S^2} \Phi \wedge \Phi= -\frac{8\pi^2}{9}\,.
\end{equation}

\end{appendix}

 \bibliography{ArXiv_v2bib}
\bibliographystyle{JHEP}

\end{document}